\documentstyle[preprint,revtex]{aps}
\begin{document}
\draft
\preprint{ETH-L and RU96-5-B}
\preprint{June 1996}
\begin{title}
The $1/r^2$ t-J model in a magnetic field\\ 
\end{title}
\author{James T. Liu$^{1}$ and D. F. Wang$^2$}
\begin{instit}
$^1$Department of Physics,
The Rockefeller University\\
1230 York Avenue, New York, NY 10021-6399, USA\\
$^2$Institut de Physique Th\'eorique,
Ecole Polytechnique F\'ed\'erale de Lausanne\\
PHB-Ecublens, CH-1015 Lausanne, Switzerland\\
\end{instit}
\begin{abstract}
We 
study the one-dimensional supersymmetric t-J model with $1/r^2$
interaction threaded by magnetic flux.  Because of the long range
interaction, the effect of this flux leads to a modification of the
electron hopping term.
We present an exact solution of this model for all values of the flux,
concisely formulated as a set of Bethe-ansatz-like equations.  
This allows a computation of the persistent currents at zero temperature.
It is found that the system violates Leggett's conjecture
despite the fact that the long range t-J model
is a special example of the Luttinger liquid universality class.

\end{abstract}
\pacs{PACS number: 71.30.+h, 05.30.-d, 74.65+n, 75.10.Jm }
\narrowtext


Exact solutions have provided us with an interesting way to deal
non-perturbatively with systems of strongly correlated electrons.
Notable examples are the electron systems with delta function
interaction\cite{yang1} and the Hubbard model \cite{wu}.  Both of
these models are solvable by Bethe-ansatz and have played an
important role in understanding the physics of the one-dimensional
electron gas.  A particularly interesting class of lattice models that are
exactly solvable despite long range interactions are the Haldane-Shastry
spin chain of $1/r^2$ exchange interaction \cite{haldane,shastry}
and its many variations.  The latter include the supersymmetric
t-J models of long range hopping and exchange
\cite{kura,kawa,wang1,wang2,wang3,hahaldane} as well as multi-component
generalizations.  Since the quasiparticle interactions in these models
are essentially two-body in nature, this fact allows a
Bethe-ansatz-like solution.  In the spirit of the Bethe-ansatz, this
indicates that the interactions are statistical in nature and arise from
demanding periodicity of the closed spin chain.

While it is natural to close the chain by imposing periodic boundary
conditions, we may also consider the case where the closed chain is threaded
by magnetic flux. Due to the non-trivial topology, electrons transported
around the chain then pick up an Aharonov-Bohm phase in the presence
of a non-zero flux.  In models with nearest neighbor exchange it is
straightforward to encode this phase (and hence the flux) by imposing
twisted boundary conditions on the wavefunctions.  However such a
prescription needs to be modified in the presence of long range interactions
where particles may hop between any arbitrary pair of sites.  We show how
this may be done for the supersymmetric t-J model with inverse square
interaction and investigate its ground state and full energy spectrum.  Based
on previous results in the absence of magnetic flux \cite{wang3}, we
derive a set of Bethe-ansatz-like equations appropriate to twisted boundary
conditions.

We consider a system of electrons on a one dimensional ring described by 
the $SU(2)$ supersymmetric t-J model.  For a uniform and closed chain of $L$
lattice sites, the Hamiltonian takes the form
\begin{eqnarray}
H_{tJ} &=&
-{1\over2}\sum_{\sigma=\uparrow,\downarrow}\> \sum_{1 \le l\ne m\le L}
[t(l-m)c_{l\sigma}^\dagger c_{m\sigma}^{\vphantom{\dagger}} + h.c.] \nonumber\\
&&+{1\over2}\sum_{1\le l\ne m \le L}\kern-10pt J(l-m)
[P_{lm}^\sigma-(1-n_l)(1-n_m)]\ ,
\label{eq:tjmodel}
\end{eqnarray}
where the hopping strength, $t(n)$, and exchange interaction, $J(n)$, are
functions only of separation due to rotational invariance of the ring.
$P_{lm}^\sigma$ is the spin exchange operator and the last term of
Eqn.~(\ref{eq:tjmodel}) encodes the hole-hole interaction with
$n_l$ the electron number operator.  We have implicitly assumed a
projection onto single occupancy at each site.

Without magnetic flux, the supersymmetric long range model has an
interaction strength given by $t(n)=J(n)= 1/d^2(n)$ where
$d(n)={L\over\pi} \sin({\pi n\over L})$.  In this case,
$d(n)$ is most readily interpreted as the chord length between sites $n$
units apart.  However, since an electron hopping along the chord length
must travel in the interior of the ring, it is not apparent how to treat the
magnetic flux with this interpretation.  Thus we use the alternate
interpretation of
$J(n)$ as the periodic version of $1/n^2$, namely
\cite{sutherland,hahaldane,fukui}
\begin{equation}
J(n)=\sum_{k=-\infty}^\infty {1\over (n+kL)^2}
=\left({\pi\over L}\right)^2{1\over \sin^2 (\pi n/L)}\ ,
\label{eq:Jn}
\end{equation}
which represents the sum of hopping over all multiples of the period $L$
with periodic boundary conditions.  It is straightforward to generalize this
instead to twisted boundary conditions appropriate to a ring threaded by
flux.  We introduce a dimensionless flux $\phi$, represented by the vector
potential $A=\phi(\phi_0/L)$ where $\phi_0 ={hc \over e}$ is the
flux quantum, so that electrons pick
up a phase $e^{2\pi i\phi}$ when transported once around the ring.  In
this case, the hopping interaction is twisted and becomes
\begin{equation}
t_\phi(n)=\sum_{k=-\infty}^\infty {e^{2\pi i\phi(n+kL)/L}\over(n+kL)^2}\ .
\end{equation}
Since the exchange interaction is insensitive to the flux, $J(n)$ is still
given by Eqn.~(\ref{eq:Jn}).  This interaction was introduced by Fukui and
Kawakami for the twisted Haldane-Shastry model in \cite{fukui} where the sum
was carried out for rational twists, $\phi=p/q$.

Previous techniques for solving the Haldane-Shastry and t-J models without
flux \cite{haldane,kura,wang3} are easily extended to the present case,
given by the twisted hopping $t_\phi(n)$.  In particular, the t-J model may
be solved by introducing a basis of Jastrow wavefunctions
describing the down-spin and hole excitations about a fully-polarized
up-spin background.  While this background may appear unnatural in the
presence of flux, it nevertheless allows an immediate generalization of the
exact solution constructed in \cite{wang3}.  All that is
necessary is to account for the twisted hopping by summing $z^Jt_\phi(n)$ 
over $n$.  This may be accomplished by using the relation
\widetext
\begin{equation}
B_2(x)\equiv {1\over\pi^2}\sum_{k=1}^\infty{\cos 2k\pi x\over k^2}
={\textstyle{1\over6}}-x(1-x), \qquad 0\le x\le 1\ ,
\end{equation}
to derive the sum formula
\begin{eqnarray}
\sum_{k=-\infty}^\infty\kern-6pt{}^\prime z^J(1-z^k)^a
{e^{2\pi i\phi k/L}\over k^2} &=&2{\pi^2\over L^2}\{
\delta_{a,0}[{\textstyle{1\over6}}(L^2-1)+\phi(1-\phi)
-(J+\phi)(L-(J+\phi))]\nonumber\\
&&\qquad+\delta_{a,1}[L-2(J+\phi)-1]
+2\delta_{a,2}\}\ ,
\label{eq:sumformula}
\end{eqnarray}
\narrowtext
\noindent
which holds for non-negative $J$ whenever
$0\le \phi \le 1$ and $0\le a \le L-1-J$ (the $'$ indicates the exclusion
of all multiples of $L$ from the sum).  This generalizes the formula for
rational twists presented in \cite{fukui}, and reduces to the standard
result \cite{haldane} when $\phi=0$.  As a result, the restriction to
rational twists is unnecessary, and the following results are applicable to
all values of the flux, indicating that there is no distinction between
rational and irrational twist angles in this strictly one-dimensional
system.  We also note that $\phi$ enters only linearly in the sum
formula, which is the reason behind the linear spectral flow found in
\cite{fukui}.

Since the three terms in Eqn.~(\ref{eq:sumformula}) correspond to
constant, 2-body, and 3-body terms in the Hamiltonian
\cite{haldane,kura}, we see that the flux has no effect on the
cancellation of 3-body terms.  This immediately shows that the
quasiparticles remain free, up to statistical interactions, even in the
presence of flux.  From the constant and 2-body terms, it is apparent that
$\phi$ acts to shift the quasiparticle momenta, leading to a modified
dispersion relation.  Since this is the extent of the modification to the
solution caused by non-zero $\phi$, the results of \cite{wang3} are easily
extended to the case of twisted hopping.  For a spin chain with
$M_\downarrow$ down spins and $Q$ holes, we introduce two sets of
pseudomomenta: $p_i$ ($i=1,2,\ldots,M_\downarrow+Q$) and $q_\alpha$
($\alpha=1,2,\ldots,Q$).
The solution to this supersymmetric t-J model may then be written
in a Bethe-ansatz-like form
\begin{eqnarray}
&&p_i L=2\pi J_i +\sum_{j\ne i}\theta(p_i-p_j)-
\sum_\alpha \theta (p_i-q_\alpha)\ ,\nonumber\\
&&\sum_i \theta (q_\alpha -p_i) = 2\pi I_\alpha\ ,
\label{eq:aba}
\end{eqnarray}
where the step function $\theta (x) =\pi\,{\rm sgn } (x)$ is the
(statistical) scattering phase.  The fermionic quantum numbers $J_i$ and
$I_\alpha$ are either integers or half-integers and are restricted to lie in
the ranges $|J_i|\le (L-M_\downarrow+1)/2$ and
$-(M_\downarrow+Q)/2\le I_\alpha\le(M_\downarrow+Q)/2-1$ respectively.
Since the $q_\alpha$'s label the hole
degrees of freedom, it gives rise to a natural splitting of the
pseudomomenta $\{p_i\}$ into $M_\downarrow$ spin and $Q$ hole degrees
of freedom.  Namely we take ${\cal Q}$ to be the set of
pseudomomenta $p_i$ satisfying
\begin{equation}
\sum_\alpha[\theta(p_i-q_\alpha)-\theta(p_{i-1}-q_\alpha)]=2\pi\ .
\end{equation}
There are exactly $Q$ such pseudomomenta, corresponding to the hole
excitations.  The remaining $M_\downarrow$ pseudomomenta then correspond
to spin excitations.  Using this distinction, the energy spectrum and
momentum of the system are given by
\begin{eqnarray}
E(\phi) &=& {\pi^2\over 6} L(1-1/L^2)
+\sum_{i\notin{\cal Q}}\epsilon_0(p_i)
+\sum_{i\in{\cal Q}}\epsilon_\phi(p_i)\ ,\nonumber\\
P(\phi)&=&(L-1)\pi+2\pi\phi (1-Q/L)-\!\!\!
\sum_{i=1}^{M_\downarrow+Q}\!(p_i-\pi)\>\hbox{mod}\>2\pi,\nonumber\\
\label{eq:micro}
\end{eqnarray}
where the single particle dispersion relation has the form
\begin{equation}
\epsilon_\phi(k)={\textstyle{1\over2}}[(k+2\pi \phi/L)^2-\pi^2
+4\pi^2\phi(1-\phi)/L^2]\ ,
\label{eq:disperse}
\end{equation}
for $0\le\phi\le1$.
This is the generalization of the results of \cite{wang3} to the case
of twisted boundary conditions.

States in the excitation spectrum are labeled by individually
non-overlapping quantum numbers $J_i$ and
$I_\alpha$.  The $J_i$ may be represented as a string of 0's and 1's
of length $L-M_\downarrow$ with $M_\downarrow+Q$ occupied positions
represented by 1's \cite{haldane2}.  The $I_i$ then in turn label
which of these 1's correspond to hole excitations (and hence are
sensitive to the flux).  The interpretation of the Bethe-ansatz-like
equations, (\ref{eq:aba}), is to separate the spin excitations by
inserting a 0 before every spin excitation.  The resulting string then
specifies the pseudomomenta $p_i$, lying in the range $[-\pi,\pi)$.
{}From Eqn.~(\ref{eq:disperse}), it is evident that states in the middle of
the string have lowest energy.  Therefore the ground state of the t-J model,
in a sector of fixed $M_\downarrow$ and $Q$, has pseudomomenta of the
general form
$p_i\in (\ldots0010101\underline{1111}0101000\ldots)$, with the hole
excitations centrally located (and underlined).  In order to study the
ground state properties, we
introduce uniform spin and hole momenta, $J_s$ and $J_h$ (integral or
half-integral as appropriate), perturbing the string of 1's to the left
or right.  In this case, the corresponding eigenenergies are
\begin{eqnarray}
{L^2\over\pi^2}E(J_s,J_h)&=&E_0 +2M_\downarrow J_s^2
+2Q[(J_h+\phi)^2\nonumber\\
&&\qquad+(J_s+J_h+\phi)^2-2\phi^2]\ ,
\label{eq:jastrowE}
\end{eqnarray}
where
\begin{eqnarray}
E_0&=&{\textstyle{1\over6}}L(L^2-1)
+{\textstyle{2\over3}}(Q+M_\downarrow)[(Q+M_\downarrow)^2-1]
+4\phi Q\nonumber\\
&&-{\textstyle{1\over2}}(Q+M_\downarrow)L^2
-{\textstyle{1\over2}} Q(Q+M_\downarrow)(Q+2M_\downarrow)\ .
\label{eq:ezero}
\end{eqnarray}
For fixed $M_\downarrow$
and $Q$, the ground state has both $J_s$ and $J_h+\phi$ as close to zero
as possible.  These states correspond to exact Jastrow product
wavefunctions describing the ground state as well as uniform excitations of
the t-J model.

To further examine the ground state of this model, we work at a fixed
hole fraction, $n_h\equiv Q/L$.  Denoting the number of
electrons by $N_e=M_\uparrow+M_\downarrow=L-Q$,
the ground state is either an $SU(2)$ singlet for even $N_e$
or a doublet for odd $N_e$.  Due to finite size effects, the ground state
properties depend on the value of $N_e\hbox{ mod }4$.  At zero flux,
whenever $N_e\ne2\hbox{ mod }4$, the ground
state carries momentum and hence is two-fold degenerate.
However this degeneracy is always lifted for non-zero $\phi$ which
breaks time reversal symmetry.  
The exact ground state energies and momenta are
given in Table~\ref{tbl:gs}, where the bulk quantities are
\begin{eqnarray}
E_g&=&-{\pi^2L\over12}[n_h(3-n_h^2)+2(3+2n_h)/L^2]\ ,\nonumber\\
P_g&=&2\pi\phi(1-n_h)\ .
\end{eqnarray}
The periodicity of the ground state
under spectral flow depends on $N_e$ and is currently under investigation;
in general the periods are fairly long and may even exceed
$L$.  We note that
the linear spectral flow apparent from the cancellation of $\phi^2$
terms in Eqn~(\ref{eq:disperse}) indicates that the ground state
at zero flux is always the absolute lowest energy state since any flow to
lower energy has nowhere to terminate.  Hence the ground state
is always diamagnetic, regardless of the number of electrons.
It follows that the persistent current of the ground state,
$I(\phi)=-{1\over2\pi}\partial E(\phi)/\partial \phi$,
always flows in one direction, regardless of
whether $N_e$ is even or odd, indicating that the well-known parity
effect for the persistent currents of the
non-interacting theory disappears.  Without flux, this long range 
t-J model belongs to the universal fixed point of one-dimensional
interacting electron systems, well-known as a Luttinger liquid in the sense
of Haldane.  The exact solutions of this model are in contrast to the 
conjecture of Leggett that the parity-effect remains valid for Luttinger
liquids\cite{legg1,legg2}.  This is apparently a result of the
long-range nature of the electron hopping.

While this model is $SU(1|2)$ invariant at zero flux, twisting the
boundary conditions on the electrons and not the holes clearly
destroys this supersymmetry.  This is also apparent from the
asymmetrical form of the Bethe-ansatz-like equations where the flux
enters only for hole degrees of freedom.  Nevertheless, it is apparent
that some form of Yangian symmetry remains \cite{haldanetalk,fukui}.

We now turn to a generalization to the $SU(1|K)$ supersymmetric t-J model
with long range interactions.  Since the two-body nature of the
quasiparticle interactions is unaffected by the flux, we may approach the
$SU(1|K)$ generalization via the asymptotic Bethe-ansatz (ABA), which was
constructed at zero flux in \cite{kawa,liu}.  Since
the magnetic flux twists all $K$ fermionic species identically,
it is natural to write the ABA in
terms of fermionic excitations above the purely bosonic vacuum
(which we denote $F^KB$).  For this choice of grading, only the
first nesting is affected by the flux.  To proceed, we introduce
$K$ sets of pseudomomenta
\begin{equation}
p_i^{(a)}:\quad i=1,2,\ldots,N_a\ ,
\end{equation}
where $a=1,2,\ldots,K$ and $N_a=\sum_{i=a}^K M_i$ ($M_i$ is the number of
electrons with spin component $i$).  Note that $N_1$ gives the total
number of $SU(K)$ electrons.  These quasi-momenta satisfy the following ABA
equations:
\widetext
\begin{eqnarray}
p_i^{(1)} L &=& 2\pi J_i + \sum_{\alpha} \theta (p_i^{(1)} - p_\alpha^{(2)})\ ,\nonumber\\
\sum_i \theta (p_\alpha^{(2)} -p_i^{(1)}) &=& 2\pi I_\alpha^{(2)} + \sum_\beta
\theta (p_\alpha^{(2)}-p_\beta^{(2)} ) 
- \sum_\gamma \theta (p_\alpha^{(2)} - p_\gamma^{(3)} )\ ,\nonumber\\
%
%
&\vdots&\nonumber\\
\sum_\gamma \theta (p_\alpha^{(K)} -p_\gamma^{(K-1)} )
&=& 2\pi I_\alpha^{(K)}
+ \sum_\beta \theta (p_\alpha^{(K)} -P_\beta^{(K)})\ .
\end{eqnarray}
\narrowtext
The quantum numbers $\{J_i\},\{I_\alpha^{(2)}\}, \ldots,
\{I_\alpha^{(K)}\}$ are integers or half-integers which are
distinct within each set respectively.  This set of equations is unchanged
from the case without flux \cite{kawa,liu}; the only place where $\phi$
enters is in the energy and momentum, given by
\begin{eqnarray}
E(\phi) &=& -{\pi^2\over 6} L(1-1/L^2)-\sum_{i=1}^{N_1}
\epsilon_\phi(p_i^{(1)})\ ,
\nonumber\\
P(\phi)&=&\sum_{i=1}^{N_1} (p_i^{(1)}+2\pi\phi/L-\pi) \ {\rm mod} \ 2\pi\ .
\end{eqnarray}
This provides the exact energy spectrum of the $SU(K)$ t-J model in the
presence of flux $\phi$, even in the non-asymptotic regime.

For the ordinary $SU(2)$ t-J model threaded by flux, the ABA
equations given in the $F^2B$ grading provide a more
symmetrical description than the microscopically derived equations,
(\ref{eq:micro}).  Nevertheless, the $BF^2$ picture of Eqn.~(\ref{eq:aba})
has an advantage in that the complete level degeneracies are
understood \cite{wang3} independent of the $SU(1|2)$ supermultiplet
structure, which is spontaneously broken for non-zero flux.

In the above, we have given an exact solution to the one-dimensional
t-J model with inverse square hopping and exchange in the presence
of an Aharonov-Bohm flux.  Since the flux breaks time reversal
symmetry, the ground state at non-zero flux is always unique.  A
consequence of the long range interactions is a modification of
the quadratic dispersion relation, Eqn.~(\ref{eq:disperse}).  The
spectral flow is piecewise linear, with cusps at integral values of $\phi$.
This further indicates that the ground state of this system is
always diamagnetic.  Thus the well-known parity effect for persistent
currents disappears for this long range t-J model belonging to
the general class of Luttinger liquids. We have also 
confirmed our conclusions by exact diagonalization on small lattices.
These results indicate a breakdown of the conjecture of
Leggett\cite{legg1,legg2} that the parity effect persists in the presence
of interactions.  It is worthwhile to investigate in further detail
the general conditions under which this breakdown may occur.
Finally, the form of the exact solution, only slightly
changed in the presence of flux, indicates that this model remains
integrable, even though the manifest $SU(1|2)$ supersymmetry has
been lost.  It remains an open issue to find an infinite set of commuting
constants of motion in the presence of magnetic flux.

\bigskip

We wish to thank C. Gruber, H. Kunz, C. A. Stafford,  
R. Khuri and X. Q. Wang for stimulating
discussions. In particular, we wish to thank C. A. Stafford
for fruitful discussions on mesoscopic systems.  
This work was supported in part
by the U.~S.~Department of Energy under grant no.~DOE-91ER40651-TASKB,
and by the Swiss National Science Foundation.

\begin{table}
\caption{Exact ground state energies $E(\phi)$ and momenta $P(\phi)$
of the t-J model with
twisted boundary conditions for $0\le\phi\le{1\over2}$.  Taking into
account a level crossing at
$\phi={1\over2}$, the absolute ground state for ${1\over2}
\le\phi\le1$ instead has energy $E(1-\phi)$ and momentum $-P(1-\phi)$.}
\begin{tabular}{cll}
$N_e\,\hbox{mod}\,4$ &  $E(\phi)$  & $P(\phi)$ \\
\hline
0&$E_g+\pi^2n_h/L$&$P_g+\pi(1+n_h)$\\
1&$E_g+\pi^2[2+n_h(1+8\phi)]/4L$&$P_g-{\pi\over2} (1-n_h-1/L)$\\
2&$E_g+4\pi^2\phi n_h/L$&$P_g$\\
3&$E_g+\pi^2[2+n_h(1+8\phi)]/4L$&$P_g-{\pi\over2} (1-n_h+1/L)$
\end{tabular}
\label{tbl:gs}
\end{table}

\end{document}